%% file: template.tex
\documentclass[journal]{vgtc}                     

\onlineid{1446}

\vgtcpapertype{Representation \& Interaction}

\title{Visualization Atlases: Explaining and Exploring Complex Topics through Data, Visualization, and Narration}

\author{%
  \authororcid{Jinrui Wang}{0009-0007-9313-5180},
  \authororcid{Xinhuan Shu}{0000-0002-9736-4454},
  \authororcid{Benjamin Bach}{0000-0002-9201-7744}, and
  \authororcid{Uta Hinrichs}{0000-0001-7494-0941}
}

\authorfooter{
  \item
  	Jinrui Wang and Uta Hinrichs are with The University of Edinburgh.
    E-mail: \href{mailto:s2251950@ed.ac.uk}{s2251950}, \href{mailto:uhinrich@ed.ac.uk}{uhinrichs}@ed.ac.uk.

  \item
  	Xinhuan Shu is with Newcastle University.
  	E-mail: \href{mailto:xinhuan.shu@newcastle.ac.uk}{xinhuan.shu@newcastle.ac.uk.}
  \item
  	Benjamin Bach is with Inria and The University of Edinburgh.
  	E-mail: \href{mailto:benjamin.bach@inria.fr}{benjamin.bach@inria.fr}}

\abstract{%
 \input{sections/abstract}
}


\graphicspath{{figs/}{figures/}{pictures/}{images/}{./}} 

\usepackage{tabu}                      
\usepackage{booktabs}                  
\usepackage{lipsum}                    
\usepackage{mwe}                       
\usepackage{array}
\usepackage{changepage}

\usepackage{mathptmx}                  

\begin{document}


\firstsection{Introduction}

\maketitle

\input{sections/introduction}

\section{Visualization Atlases: Examples and Collection}
\label{sec:collection}
\input{sections/examples}

\section{Visualization Atlas Design Patterns}
\label{section:survey}
\input{sections/dimensions}

\section{Interviews with Visualization Atlas Creators}
\label{sec:interviews}

\input{sections/interview}

\input{sections/genres}

\input{sections/characteristics}

\input{sections/discussion}

\acknowledgments{%
We thank our reviewers for their invaluable insights and critiques that greatly enhanced our work. 
Special appreciation goes to the following atlas creators for their participation and unique perspectives in interviews: Dominikus Baur, Attila Bátorfy, Tommaso Elli, Alenka Guček, Rob Korzan, M.Besher Massri, Alice Thudt, Nil Tuzcu (listed in alphabetical order by surname).
}

\bibliographystyle{abbrv-doi}

\bibliography{template}
\appendix 

\end{document}

%% file: sections/abstract.tex
This paper defines, analyzes, and discusses the emerging genre of \textit{visualization atlases}. We currently witness an increase in web-based, data-driven initiatives that call themselves ``atlases'' while explaining complex, contemporary issues through data and visualizations: climate change, sustainability, AI, or cultural discoveries. 
To understand this emerging genre and inform their design, study, and authoring support,  
we conducted a systematic analysis of 33 visualization atlases and semi-structured interviews with eight visualization atlas creators. Based on our results, we contribute (1) a definition of a visualization atlas as a compendium of (web) pages aimed at explaining and supporting exploration of data about a dedicated topic through data, visualizations and narration. 
(2) a set of design patterns of 8 design dimensions, (3) insights into the atlas creation from interviews and (4) the definition of 5 visualization atlas genres.
We found that visualization atlases are unique in the way they combine i) exploratory visualization, ii) narrative elements from data-driven storytelling and iii) structured navigation mechanisms. They  target a wide range of audiences with different levels of domain knowledge, acting as tools for study, communication, and discovery. We conclude with a discussion of current design practices and emerging questions around the ethics and potential real-world impact of visualization atlases, aimed to inform the design and study of visualization atlases.

%% file: sections/introduction.tex
Over the past years, we noticed an increase in highly polished web-based projects that use visualizations to make data about complicated contemporary problems accessible to a wide audience. Projects like \textit{The Atlas of Economic Complexity}~\cite{GrowthLab2013Atlas}, \textit{Our World in Data}~\cite{RoserOurWorldInData}, or \textit{The Atlas of Sustainable Development Goals}~\cite{Pirlea2023AtlasSDG} summarize and explain large and multidimensional datasets and explain key insights about global challenges, such as climate change, sustainable development, economics, artificial intelligence, or cultural discoveries. Many of these projects are created by large organizations such as the World Bank and the UN, or research labs at prestigious institutions, with a mission to promote open data, methodological transparency, and education. 

Many of these projects call themselves ``atlas'', ``observatory'', or ``index'' and cannot easily be described through existing genres or forms of visualization alone, such as dashboards~\cite{bach2022dashboard}, narrative visualizations~\cite{segel2010narrative}, collections of journalistic articles~\cite{hao2024design}, or interactive  applications. In fact, these projects seem to act as umbrellas that cover a palette of techniques, providing different perspectives and tools to investigate, explore, and understand a topic through its data: interactive visualizations, key performance indicators, overviews, details-on-demand, narration and storytelling---all of which are embedded into well-organized pages, offering bespoke navigation aids, overview pages with introductions, and mechanisms for onboarding. 
However, we lack a detailed understanding of the uniqueness and characteristics of these kind of projects, which we decided to term \textit{visualization atlases}, motivated by the fact that \textit{a)} `atlas' is the term most frequently used across projects and \textit{b)} its compelling analogy to geographic atlases with their purpose of mapping `unknown' terrain for lookup, learning, and exploration. 
Hence, we ask \textit{What is a visualization atlas and how to inform its design?} We aim for a definition, design patterns, a typology of atlas types, and to understand the motivations that lead to their creation. 
We believe this can inform future atlases, further study, and respective authoring tools.

To that end, this paper adopts two complementary methods. First, we collected 33 atlases from the internet, as no scientific writings on visualization atlases exist to date. We then analyzed those atlases' design and identified 45 \textbf{design patterns} along 8 design dimensions (\cref{section:survey}): \textit{content page design}, \textit{entry page design}, \textit{atlas structure}, \textit{visualization interaction}, \textit{atlas navigation}, \textit{visualization types}, \textit{atlas onboarding}, and \textit{data transparency}. Second, we conducted \textbf{semi-structured interviews} with eight atlas designers (\cref{sec:interviews}) who specialized in visualization design. We asked about the motivation for the atlas project, the intended audiences, and what they see as the characteristics of this atlas format. These two approaches allowed us to identify 5 higher-level visualization atlas \textbf{genres} (\cref{section:genres}) that describe usage scenarios and imply design patterns. 

Finally, we formulate 9 key characteristics of atlases with respect to three main aspects: \textsc{Topic} (Visualization atlases present \textit{complex} topics to a wide range of audiences in a \textit{data-driven} and \textit{comprehensive} way), 
\textsc{Curation} (Visualization atlases are \textit{scoped}, \textit{structured}, and \textit{visually curated}), and \textsc{Visualization} (Visualization atlases are \textit{visualization-driven}, \textit{explanatory} and
\textit{exploratory}) (\cref{sec:characteristics}).
Based on these characteristics, we formulate a shorter definition of visualization atlases that is more memorable: it defines a visualization atlas as a compendium of (web) pages aimed at explaining and supporting exploration of data about a dedicated topic through data, visualizations and narration.

All materials including atlas cases and detailed descriptions of design patterns can be found online: \url{https://vis-atlas.github.io}.

%% file: sections/examples.tex
\begin{figure*}[!t]
  \centering
  \includegraphics[width=1\textwidth]{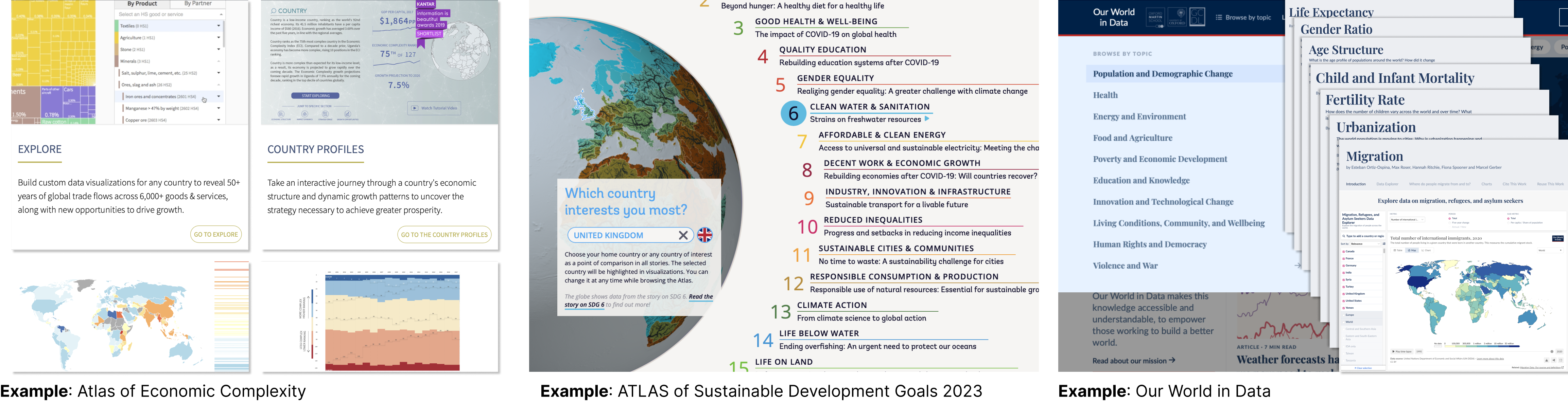} 
  \caption{Three visualization atlas examples.}
  \label{fig:atlas-examples}
  \vspace{-1.5em}
\end{figure*}

We start our investigation by describing three representative examples of visualization atlases~ (\cref{fig:atlas-examples}) to give background and prepare our more detailed investigations in Sections \ref{section:survey} and \ref{sec:interviews}.

\textbf{The Atlas of Economic Complexity}~\cite{GrowthLab2013Atlas} ({\textsc{EconComplexity}) by Harvard's Growth Lab reports on economic dynamics and growth opportunities for 250~countries worldwide, centering on the Economic Complexity Index (ECI). The online visualization atlas is fed by research published in an academic book of the same name~\cite{Hausmann2013} but supports interactive exploration and storytelling to present the underpinning economic theories. The atlas is designed for long-term usage; the underlying data is updated on a yearly basis. 

The atlas features four qualitatively very different complementary components, shown on its landing page in~\cref{fig:atlas-examples}-2. The first component is an exploratory visualization tool where a visitor can freely customize the data dimensions and countries they are interested in. A variety of visualizations are provided in full screen: treemaps, node-link diagrams, stream graphs, maps, etc. A step-by-step walkthrough familiarizes users with the customizing features. The second component is a set of dedicated country profiles; each profile starts with a key metrics overview and then features a slideshow \cite{segel2010narrative} explaining how economic complexity theory applies to a country. Readers can move between slides linearly or choose topics from a menu including market dynamics, growth opportunities, export statistics, etc. On each slide, the visualization is almost full-screen and accompanied by factual and automatically-generated statements (\textit{``The United States of America exported products worth USD \$2.55 trillion in 2021''}). The remaining two components in this atlas are an article that explains growth projection methods and a page featuring three interactive visualizations for reporting country \& product complexity rankings. The atlas features extensive documentation on methodology and data sources; videos are provided for atlas usage, key concepts, research talks, and publications. 

The atlas is used in Harvard classes and the Growth Lab's researchers use it for field research with governments across the world. Policy makers in local and national governments use it as a reference tool. The atlas is supported by a dedicated team of designers and developers based at the Growth Lab which is authoring additional visualizations and storytelling pieces, such as another visualization atlas for economic data on a city level (\textsc{Metroverse}~\cite{tuzcu2023unraveling}).

\textbf{The Atlas of Sustainable Development Goals 2023} \cite{Pirlea2023AtlasSDG} (\textsc{SDG2023}) shows the lastest insights into global progress and challenges towards achieving the 17 SDGs~\cite{united2023sustainable} defined by the United Nations in 2015.
The project was commissioned by the World Bank as a sequel to their previous atlases in 2017~\cite{WorldBank2017SDGAtlas}, 2018~\cite{WorldBank2018SDGAtlas}, and 2020~\cite{Pirlea2020SDGAtlas}. Data are mainly from the World Development Indicator compiled by the World Bank about global development and living quality. 
\textsc{SDG2023} consists of 17 pages, each dedicated to an individual goal and starting with a brief description of the respective goal and its sub-targets.  
Visualizations include a huge variety of well-designed linecharts, treemaps, small multiples, and others. Many of the visualizations are explained through scrollytelling, using piece-meal explanations and animations. 

\textbf{Our World in Data} (\textsc{OurWorldInData})~\cite{RoserOurWorldInData} aims at addressing global challenges using data-driven methods. While not explicitly calling itself an atlas, the project shares significant similarity with the other two examples: it publishes diverse data about poverty, hunger, diseases, under an extensive topic hierarchy (10 topics, 115 subtopics); it makes consistent use of data visualizations; it provides in-depth analysis written by scientists; and it aims to make data and knowledge public. As of March 2024, the atlas features over 4543 charts. Rather than relying on a limited set of data sources, part of the mission of \textsc{OurWorldInData} is to find and aggregate diverse data sources and republish them for dissemination. 

%% file: sections/dimensions.tex
To inform our investigation of atlas design and characteristics, we systematically collected atlas examples and identified respective design patterns. Here, we report on the collection and consequent coding.

\begin{figure*}[tb]
  \centering
  \includegraphics[width=1\textwidth]{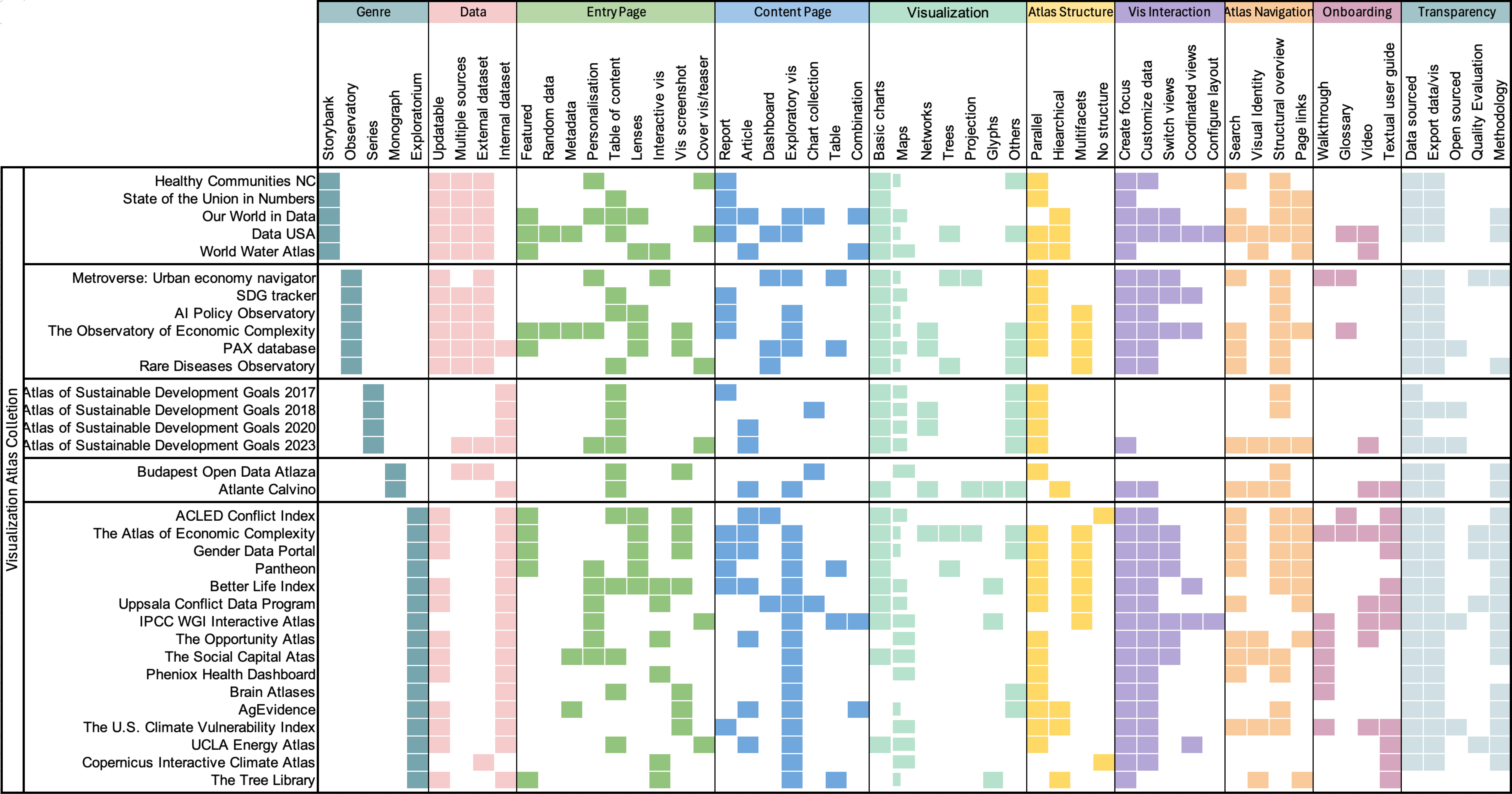} 
  \caption{Complete coding of 33 visualization atlases, including eight design dimensions plus two contextual dimensions of data, and genres. Differently sized squares in the `visualization' dimensions indicate prominence of a visualization across the atlas.}
  \label{fig:coding}
  \vspace{-2em}
\end{figure*}

\subsection{Structured Atlas Collection and Design Classification}

\textbf{Phase 1: Seed collection}---We started from a seed collection of 8 renowned or award-winning atlases we came across on the internet that focus on global issues, three of which are described in the above section. We found they are usually named as ``atlas'', ``observatory'', ``index'', or ``portal'' featuring complicated datasets.

\textbf{Phase 2: Keyword search}---Based on the seed collection, we considered two branches of collection strategies: by project names and by creators. We began the keyword search by using  common names found in the seed collection like ``visualization atlases, ``data observatories'', and kept those focusing on specific domain topics. 
Meanwhile, we searched related projects created by global data publishers like universities, IGOs and their collaborating design agencies.
The collection was further expanded when we interviewed creators who contributed cases from their related fields. This process led to a pool of 41 cases.

\textbf{Phase 3: Refine}---We further examined all the cases and iteratively refined our understanding of visualization atlases. We delineated its scope by considering three exclusion criteria.
First, we prioritized web-based atlases in this paper and excluded the printed forms, due to our deliberate focus on open data practices and accessibility to datasets. 
These printed atlases (e.g., ~\cite{Atamny2017, LateralNorth2019, CheshireUberti2021}) were limited to affording interactions and exploration. 
Second, we discarded some cases that were purely online repositories for datasets (e.g.,~\cite{ GoogleArtsCultureExplore}), as we considered visualizations a pivotal component in atlases for exploration and explanation. 
Third, we also carefully discarded cases  lacking an observable structural curation or those ambiguous in selecting a definitive topic (e.g.,~\cite{ReutersGraphics}).
As a result, we reached a collection of 33 atlases.

\textbf{Phase 4: Pattern classification}---We went through several iterations for the dimensions (pattern groups) and patterns.
Firstly, we explored the contextual background of the collection by coding each example's domain, organizational creator(s), type of organization and data sources (see supplementary materials). Then, we analyzed the design composition of atlases to explore their design space. We looked at \textit{a)} concepts we knew from other visualization systems such as visualization types, means for storytelling, and onboarding mechanisms, as well as \textit{b) concepts we found were very specific to atlases such as overview pages, page structure, and navigation.
To ensure these dimensions were distinct and complete,} all authors independently coded five atlases along all those dimensions. We then resolved ambiguities and, refined, and added dimensions and patterns. For example we discarded the dimension `exploration' and merged `explanation features' with `content page styles'. For each dimension, we listed possible patterns with name and detailed description, e.g., the full screen choropleth map with filters are exploratory vis page in \textsc{Opportunity}~\cite{OpportunityAtlas2023}. Then, all authors coded the same five atlases again with the refined schemes and pattern descriptions. After several iterations, two authors coded the entire collection and eventually reached agreement (\cref{fig:coding}), including 8 dimensions and 45 patterns explained in the following sections. \\

From our classification, we describe a visualization atlas as made out of \textit{pages}, a notion that roughly corresponds to a webpage. Links on one page can lead to another. Across atlases, we identified two types of pages---content pages and entry pages--- that serve different purposes.

\subsection{Content Page and Content Page Design Patterns}
\label{sec:contentpage}

\begin{figure*}[t]
  \centering
  \includegraphics[width=1\textwidth]{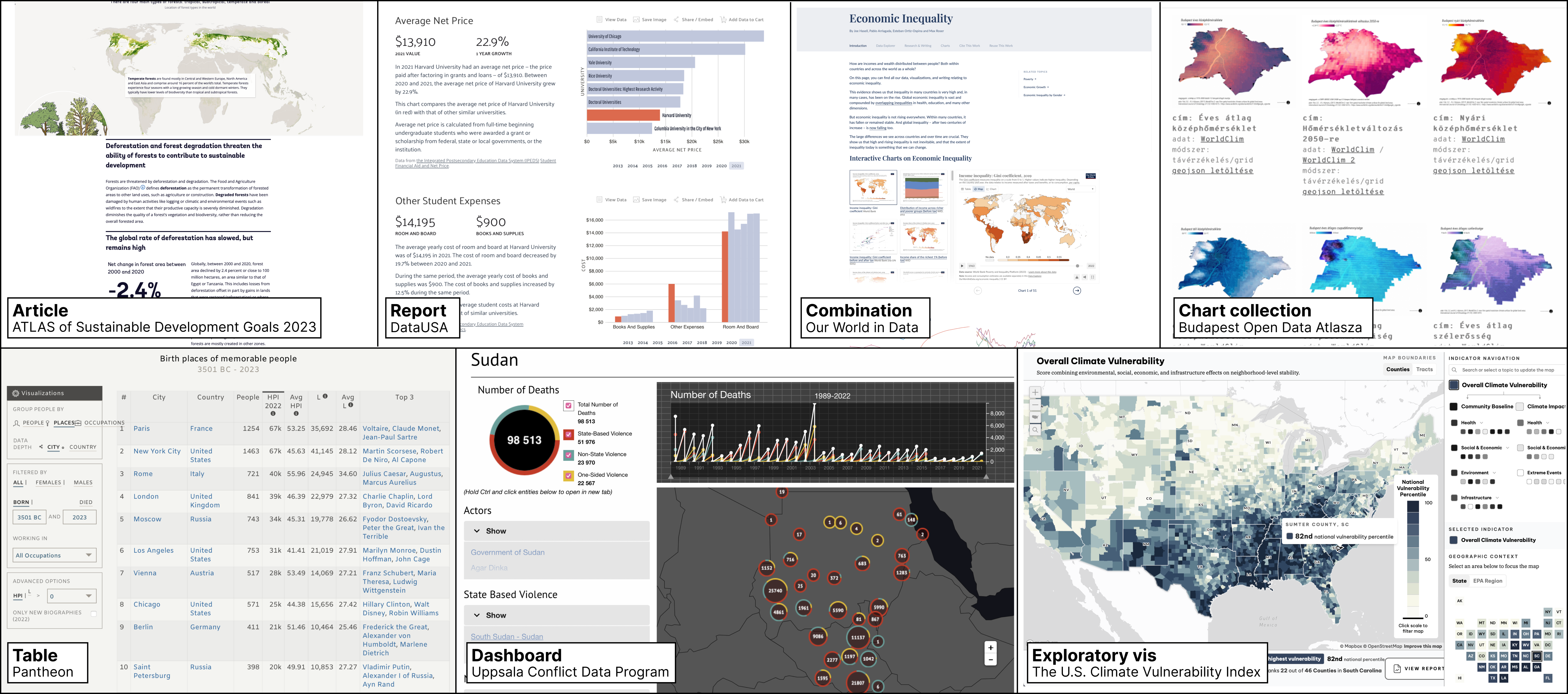} 
  \caption{Examples of content pages styles.}
  \label{fig:page-type}
  \vspace{-2em}
\end{figure*}

A visualization atlas \textbf{content page} (\cref{fig:page-type}) can include visualizations, textual descriptions, titles, author, illustrations, links to further information, etc. As centerpieces in every atlas, content pages not only provide access and allow people to explore the data presented, but also explanations. Content pages usually provide a curated view on a specific aspect within the atlas and can come in different \textit{styles}:

\begin{itemize}[noitemsep,leftmargin=*]
    \item An \textbf{article} page (12; 37\%) is a comprehensive piece of intellectual analysis about a given topic. Articles usually describe context; in-depth analysis with static or interactive visualizations to underline their argument. Articles are usually created by dedicated analysts and experts, individually or collaboratively (e.g., \textsc{SDG2023}). 

    \item A \textbf{report} page (13; 40\%) is a highly structured presentation of key insights from a dataset, with visualizations and shorter descriptions as explanations. Usually, reports are automatically created from a template with scripted text templates such as  \textit{``As of 2021, xx\% of xxx residents were US citizens''}(\textsc{DataUSA}~\cite{DatawheelUSA}). Once specified, reports can be generated for different data sets such as countries (\textsc{BetterLife}~\cite{OECD}) or topics (\textsc{DataUSA}). 

    \item \textbf{Dashboard} (6; 19\%) content pages show data in a condensed way as common in dashboards~\cite{bach2022dashboard} and provide little to no textual explanations. Usually, dashboards are also automatically created, or make it easy to create templates and populate them with different data similar to reports. An example for a dashboard is found in \textsc{UCDP} country view~\cite{UCDP}. Dashboards can be good means to imply and inform monitoring tasks where data update on a regular or irregular basis. 

    \item \textbf{Exploratory visualization}~(22; 67\%) can make an entire content page, providing a rich and interactive experience for data exploration. Many exploratory visualizations come with tooltips, and pan\&zoom as well as features for selecting data items, data dimensions, filter data, and styling options (\textsc{USClimate}~\cite{USClimateVulnerabilityIndex}). Compared with the other page styles, exploratory visualizations give users maximal agency over exploring and investigating the data on their own.

    \item A \textbf{chart collection} (4; 13\%) content page list a set of static and non-interactive charts. In the examples we found, notably in \textsc{SDG series} 2018 edition~\cite{WorldBank2018SDGAtlas}, those chart collections came without additional explanations apart from titles, brief captions and sparse annotations within the charts. Chart collections can be an easy means to publish existing charts, avoiding the manual labor of programming dashboards and reports or writing articles. They might also be a good fit for atlases that require frequent updates of standardized charts and are organized as an indexing aid.

    \item \textbf{Table} (5; 16\%) content pages list plain data values without any visualization, annotations, or explanations. As in \textsc{Pantheon}~\cite{DatawheelPantheon}, a table page can embed original data in a straightforward way.
    
\end{itemize}

While we mostly found clear cases of a content page being one of these styles, some atlases contained \textbf{combinations} (4; 13\%) of multiple page styles, without one style being dominant. In those cases, the page either contain several clearly distinguished parts, such as in \textsc{OurWorldInData} which has pages that feature sections on reports, exploratory vis, chart collections, etc.

\subsection{Entry Page and Entry Page Design}
\label{sec:entrypage}
An \textbf{entry page} design provide access to different content pages. Entry pages are essential to visualization atlases as they can provide an overview to a wider topic (or subtopic) in the atlas, deliver respective explanations and content, and, most importantly, guide the navigation to specific content pages. All atlases in our collection contain at least one entry page---the atlas landing page. Three~atlases (10\%) contain entry pages for specific subtopics in the atlas (e.g., the city overviews in \textsc{Metroverse}). Unlike content pages that have specific styles, entry pages use combinations of textual and visual interactive \textit{components}, discussed below.

A first group of entry page components provides links and access to content pages. For example, a \textbf{table of contents} (17; 52\%) shows the atlas structure. Other entry pages introduce different exploration angles and a mixture of features through what we call \textbf{complementary lenses} (10; 31\%), such as the lenses presented in \textsc{GenderData}~\cite{WorldBankGender}.
While some atlases show \textbf{featured pages} (10; 31\%) to point to selected content pages and aim to draw visitors' attention to the latest or highlighted pages (\textsc{OurWorldInData}). In two of the cases (7\%), \textbf{random data} links take users to a random content page of a particular subtopic, which may trigger serendipitous findings (\textsc{Pantheon}). 

A second group of entry points aim to provide previews on data, by showing a statistic summary or visualization. \textbf{Metadata} (4; 13\%) demonstrate the amount and diversity of data available in the atlas, such as the number of pages in a subtopic~(\textsc{DataUSA}). For those that use visualizations for data preview, we identify \textbf{vis screenshots} (10; 31\%) that cropped directly from the content pages (\textsc{EconComplexity}, \cref{fig:atlas-examples}); and \textbf{cover vis} (6; 19\%) that is a teaser specifically designed for the entry page, e.g., the rotating globe in \textsc{SDG2023} (\cref{fig:atlas-examples}).

A third and last group of components on entry pages aims to foster engagement. \textbf{Exploratory vis} (8; 25\%) can serve directly as an entry page, e.g., an interactive choropleth map page that links to individual pages (\textsc{Opportunity}). Some entry pages allow users to \textbf{personalize} (11; 34\%) their experience and entry point by, enabling the selection of a country of interest (\textsc{SDG2023}) or specifying personal priorities on a set of sliders (\textsc{BetterLife}).

\subsection{Atlas Structure Design Patterns}
\label{sec:structure}

The atlas structure describes how content pages are organized and linked. The page organization can combine different structural options; 12 atlases (37\%) in our collection combine two structures. 

\begin{itemize}[noitemsep,leftmargin=*]

\item In a \textbf{parallel structure} (26; 79\%), content pages have the same page styles (\cref{sec:contentpage}) and roughly follow the same structure and elements inside each content page. In \textsc{SDG2023}, all pages are articles, use scrolly-telling, show similar information about each SDG, and follow the same layout, colors, and other design elements. \textsc{USClimate} is an example that creates reports for each location, based on exactly the same report template. Parallel structures can make it easy for visitors to create a mental model of the atlas (\cref{sec:onboarding}, \cref{sec:navigation}) because of repeated and familiar designs. Parallel structures can also help by adding more data and pages in the future. Parallel structures also suggest that the atlas topic is presented from the perspectives of several `main characters', i.e., individual concepts such as indicators, countries, persons, etc.

\item In a \textbf{hierarchical structure} (7; 22\%), content pages are grouped into \textit{subtopics}. Usually, each step of the hierarchy / subtopic is introduced by a dedicated entry page. Content pages in hierarchical structures can follow the same page style or be entirely different. In \textsc{OurWorldInData}, all content pages are organized around subtopics such as life expectancy, gender ratios, and migration, and each topic has a group of pages like articles and exploratory visualization. Another example is \textsc{Calvino}~\cite{AtlanteCalvinoProject}, where the hierarchy is illustrated on the entry page as a map with three itineraries and three stages.

\item A \textbf{multiple facets} (10; 31\%) structure is similar to parallel hierarchical structures but instead of each content page focusing on a \textit{different} part of the data, content pages represent facets ``of the same coin''. An example for multiple facets is \textsc{BetterLife}, which offers content pages about overall indicators, topic breakdowns, country comparisons and user stories. Multiple facets  provide complementary views. They can be good solutions for large and multidimensional datasets where simple parallel or hierarchical structures are too constraining. On the flip side, multiple facets might require more advanced navigation features in the atlas (see \cref{sec:navigation}).
    \end{itemize}

If an atlas consists of pages that are only loosely connected, as in \textsc{ACLED}~\cite{ACLED} that positions different conflict data resources like country dashboards and case studies without a clear categorization, then we categorize it as \textbf{no structure} (2; 7\%).

\subsection{Atlas Navigation Design Patterns}
\label{sec:navigation}

Navigation across an atlas guides the user to individual pages, helps orient which page they are currently seeing, and may suggest which parts of the atlas they have not yet seen. Navigation within an atlas and across content pages is a prime challenge in atlas design that naturally becomes more tricky the more extensive the atlas is: the more content pages, topics, deeper hierarchy and/or multiple facets.
\textbf{Search} (18; 55\%) is a simple navigation means for large atlases with complex data (\textsc{Metroverse}). Creating different \textbf{visual identities} (8; 25\%) per page or subtopic can tell a visitor where they are, e.g., by using colors or icons (\textsc{SDG2023}). \textbf{Structural overviews} (24; 73\%) include tables-of-content, menus, breadcrumbs, progress bars, and navigation-tabs that show the entire or partial page structures. Eventually, \textbf{page links} (14; 43\%) directly link to other pages within the atlas (\textsc{OurWorldInData}).

\subsection{Visualization Types Design Patterns}
\label{sec:vis}

Many visualization atlases include multiple visualization types with different levels of complexity and familiarity. 
The majority of atlases include \textbf{basic charts} (25; 76\%) (linecharts, barcharts, and variations thereof) and \textbf{maps} (29; 88\%). 
However, maps are used to different degrees: 
mainly maps; alongside other charts; and few maps (shown as differently sized rectangles in \cref{fig:coding}). 
Only 5 (16\%) atlases rely entirely on maps as the only visualization type.
Six (19\%) atlases use some sort of  \textbf{network} visualization (e.g., node-link diagram, adjacency matrix, sankey diagram). 
Other atlases include \textbf{treemaps} (5, 16\%),  \textbf{projections} (3, 10\%) and/or \textbf{glyphs} (4, 13\%) for high-dimensional data. Visualization types that occur only once or twice in the collection, like radar charts and violin charts, were coded as \textbf{other}.

\subsection{Visualization Interaction Design Patterns}
\label{sec:interaction}

This dimension describes interaction within a content page to explore data and visualizations.
\textbf{Creating focus} (29; 88\%) draws attention to or highlights selected data points through, e.g., asking which country is of users' interests (\textsc{SDG2023}).
\textbf{Switching views} (11; 34\%) changes the visualization techniques, e.g., from a treemap to geographic map (\textsc{EconComplexity}) , offering multiple complementary perspectives of the data. \textbf{Coordinated views} (6; 19\%) help compare two or more data sets; e.g., reading university reports side by side (\textsc{DataUSA}). Some atlases offer to \textbf{customize data} (25; 76\%) through filtering, brushing, selecting indicators, and visualizing different datasets such as in the example of \textsc{Opportunity}'s advanced tool. Customized views are often intrinsic in exploratory visualization page styles (\cref{sec:contentpage}).

\subsection{Data Transparency Design Patterns}
\label{sec:transparency}

Patterns in this dimension describe how an atlas' data, analysis, and visualizations are made transparent. It can be done by disclosing \textbf{data sources} (31; 94\%) through inline or endnote citations or directing to databases and respective websites; 
by \textbf{exports} (29; 88\%) to download data or charts as figures; 
through \textbf{open source} code (5; 16\%) for data analysis (\textsc{IPCCAtlas}) or visualization implementations (\textsc{SDG2023}). 
\textbf{Quality evaluations} (5; 16\%) are explicit explanations of missing data, their reliability, and possible data loss. This can happen through dedicated scores (e.g., low/medium/high data quality per city \textsc{Metroverse}) or through bespoke visualizations showing data availability (\textsc{GenderData}).
Lastly, \textbf{methodology} (20; 61\%) explanations describe how data is collected and calculated, the criteria for inclusions and exclusion, and if there exist domain research publications.

\subsection{Onboarding Design Patterns}
\label{sec:onboarding}

Our last dimension describes help resources for visitors to understand how to use an atlas in terms of contextual background, general features, and reading visualizations.
Drawing from Stoiber’s taxonomy of onboarding types for visual analytics systems~\cite{stoiber2019visualization}, we identified three similar types in atlas onboarding: step-by-step \textbf{walkthrough} (8; 25\%) tutorials that guide users through an atlas' interactive features; \textbf{introductory videos} (9; 28\%) demonstrating an atlas' usage (\textsc{IPCCAtlas}~\cite{IPCC}), structure (\textsc{Calvino}), or key findings (\textsc{SDG2023}); as well as as \textbf{textual user guides} (11; 34\%). In addition to this taxonomy, we found a new onboarding category used in five atlas cases: \textbf{glossaries} (5; 16\%) with explanations about an atlas' background, key metrics, and calculations. 

%% file: sections/interview.tex
Complementing our analysis of design patterns, we interviewed atlas creators to learn about the intended purposes of the atlases, overall motivation, design practices and challenges encountered.

\subsection{Interview Procedure}
\textbf{Participants:}
We interviewed eight visualization experts (atlas \textit{creators}) who were directly involved in the design of seven visualization atlases. Creators' professional backgrounds included 5~visualization designers, 1~full-stack vis engineer~(P4), and 1~data journalist~(P6). Although some participants have worked on several visualization atlas projects, we intentionally interviewed them based on a specific atlas. Atlas projects focused on during interviews included a range of topics (e.g., literature, sustainable development, and technology), different domains and author organizations (e.g., universities, IGOs, design agencies, and research institutes). 3~participants were internal to the atlas' author organization; 5~were hired as external designers (Table~\ref{tab:interview_list}).

\begin{table*}[ht]
\footnotesize
    \centering
    \begin{tabular}{p{2.6cm}llp{3.5cm}}
    \hline
        \textbf{Participant \& Role} & \textbf{Vis Atlas Name} & \textbf{Domain} & \textbf{Organization}\\
    \hline
        P1 - Vis designer & Atlante Calvino (\textsc{Calvino})~\cite{AtlanteCalvinoProject} &  Literature criticism & \begin{tabular}{@{}p{3.5cm}@{}}University of Geneva,\\ the Politecnico di Milano\end{tabular}\\
    \hline
        P2, P3 - Vis designers & ATLAS of Sustainable Development Goals 2023 (\textsc{SDG2023})~\cite{Pirlea2023AtlasSDG} & Sustainable development & World Bank \\
    \hline
        P4 - Full-stack engineer & OECD AI Policy Observatory (\textsc{AIPolicy})~\cite{OECDAI} & Artificial intelligence & OECD \\
    \hline
        P5 - Vis designer & Rare Diseases Observatory (\textsc{RareDisease})~\cite{RareDiseasesSlovenia} & Health & The Jožef Stefan Institute \\
    \hline
        P6 - Data journalist & Budapest Open Data Atlasza (\textsc{Budapest})~\cite{Batorfy2023Budapest} & Urban development & ATLO \\
    \hline
        P7 - Vis designer & The U.S. Climate Vulnerability Index (\textsc{USClimate})~\cite{USClimateVulnerabilityIndex} & Climate & \begin{tabular}{@{}p{3.5cm}@{}} Darkhorse Analytics, EDF,\\ Texas A\&M University\end{tabular} \\
    \hline
        P8 - Vis/UX designer & Atlas of Economic Complexity (\textsc{EconComplexity})~\cite{GrowthLab2013Atlas} & Economics & Harvard Growth Lab\\
    \hline
    \end{tabular}
    \caption{Background of eight participants from seven projects, including their roles, associated projects, project domains, and initiating organizations.}
    \label{tab:interview_list}
    \vspace{-2em}
\end{table*}

\textbf{Interview Structure:}
One-hour interviews were conducted individually with each participant, except for P2 \& P3 who were interviewed together because they worked on the same visualization atlas. The semi-structured interview questions focused on seven aspects: the visualization atlas background, goals \& motivations, collaborative process, target audience, design approaches, and a reflection on the nature of visualization atlases, compared to other visualization forms.

\textbf{Data Collection \& Analysis:}
All interviews were audio-recorded and transcribed to enable an in-depth thematic analysis~\cite{braun2012thematic}. The seven questions from the interviews directly informed our coding; for each of the questions we created a coding category using NVivo~\cite{NVivo14}. Then, we compared the coded transcripts for each category across the interviews for common sub-themes, e.g., 5 out of the 8 interviewed mentioned \textit{data insights} as a goal, forming the basis for the following findings.

\subsection{Motivation \& Goals for Creating Visualization Atlases}
\label{sec:interview-goal}

\textbf{Promote Selected Datasets for Exploration \& for Taking Action.} 
4/8~creators highlighted the promotion of a selected dataset or range of data that had been collected and curated by an associated organization as one of the core motivations for creating a visualization atlas. To that end, interactive visualizations are created to make this data explorable by people within and outside of the organization. For example, \textsc{EconComplexity}:  ``\textit{started as a side product of a book and [the Growth Lab] hoped to make the data visualizations more interactive, exploratory for people.}''~[P8]. Similarly, the \textsc{SDG series} showcased and communicated data owned by the World Bank in the form of atlases since the 70s, first in print, then digitally as PDFs and now in the form of webpages that feature interactive visualizations.

Some atlas projects aim to act as a call for action. For example, \textsc{Budapest} is based on public data about Budapest painstakingly put together by a data journalist~(P6). The atlas can be seen as a call for action for making recent national development data open access, to raise awareness on urban development issues, and to hold public administration accountable for data transparency. Similarly, the driving force of \textsc{USClimate}, is to \textit{``inform local communities and give them some advocacy power to apply for funding in more appropriate ways''}~[P7].

\textbf{Publish Insights \& Stories.}
5/8~creators reported the dissemination of insights in the form of stories and \textit{``demonstrations''} ~[P8] to be a central goal of atlas projects. For example, \textsc{Calvino} was driven by the idea of ``\textit{producing analytical narrations using visualization in literature criticism}''~[P1]. The atlas provides three narrative strands that combine literature analysis with interactive visualizations. The \textsc{SDG2023} emphasizes latest insights ``\textit{on each [sustainability] goal through a lens of current events}''~[P2]. 
Interestingly, in 3/7 projects, the goal of highlighting insights and stories arose \textit{during} the atlases' design process. For example, while designing \textsc{EconComplexity} the creators' team ``\textit{realized that people just simply do not understand [the economic complexity theory]}''~[P8], so they included narrations in the country profiles that explain the terminologies used.

\textbf{Promote Topics in a Systematic, Data-Driven Way.}
For 4/7 visualization atlases, the data was in place from the beginning, directly informing design processes. In the remaining three atlas projects, the systematic identification and curation of data sources on the topic in focus motivated the atlas project. For example, the goal of \textsc{AIPolicy}~\cite{OECDAI} was to ``\textit{build insights that are not visible from any single data sources over time}''~[P4]. The team therefore not only tried to identify data sources relevant to this specific atlas, but also came up with updatable and reproducible data sourcing methods that can be applied across different topics and atlases: ``\textit{[F]or each observatory there are generic data sets that work on any topics like [fetching latest] research papers [on this topic].}''~[P4]. Customized data sources are then included ``\textit{that are specific to the topic, [...], such as climate data for climate-related atlases, or data from Stack Overflow for tech-related atlases}''~[P4]. 

Keeping data up-to-date is key for some atlases. The team of \textsc{RareDisease}~\cite{RareDiseasesSlovenia} considered the accessibility, update frequency, and data source reliability to ensure data tracking and updates and that the visualization atlas remains relevant in the long run. They even collaborate ``\textit{with users [in] that we constantly add new data strings}''~[P5].  

\subsection{Intended Audiences of Visualization Atlases}
\label{sec:interview-audience}
Closely linked to their goals, many visualization atlases mention a wide range of target audiences, including researchers, policy makers, the media, and the general public. We interviewed creators about the intended audiences and how this influenced their design process. 

\textbf{Internal \& External Experts.}
5/7 atlases target expert audiences internal to the atlas team or external domain experts. \textit{Internal experts} work at the institution issuing the visualization atlas and are often closely involved in its design process (e.g., by informing data stories, as in the case of \textsc{SDG2023} and \textsc{EconComplexity}). They have extensive domain knowledge and are familiar with the underlying data. While their visualization expertise may vary, their involvement in the creation process grants them the ability to inform visualizations and related atlas features.
Internal experts use visualization atlases to inform research on the topics presented (\textsc{EconComplexity} \& \textsc{AIPolicy}), make presentations and/or demonstrations.  
Visualization atlases may also be used for teaching: ``\textit{[The PI] would request some of the features [...] in order to teach a certain topic in his class}''~[P1; \textsc{Calvino}].

\textit{External experts} are academics and analysts external to the institution or project team issuing the visualization atlas. They have extensive domain knowledge, but may not be as familiar with the presented data and included visualizations.
They typically will use the visualization atlas as a reference tool (e.g., in the case of \textsc{Calvino}), or to inform their research: e.g., \textsc{RareDisease} also targets doctors and clinicians, who are ``\textit{working on developing new treatments for rare diseases}''~[P5]. 

\textbf{Policy makers.}
3/7 atlas projects specifically target ``policymakers'' which can include a range of roles and usage scenarios.
In the case of \textsc{USClimate}, they are ``\textit{US-based decision makers at different levels [... who] aim to make positive systematic changes}''~[P7]. In \textsc{EconComplexity}, a pilot user study identified policy makers as ``\textit{mainly people working in the data office, both locally and nationally, [who] care about specific issues}''~[P8]. In \textsc{AIPolicy} policy makers were defined as ``\textit{not technical experts in AI, but [people who] would like to be informed about AI progress, especially in their own countries, and other similar countries''}~[P4]. Our creators agreed that senior policy makers, depending on their background, expertise, and interest, may not directly engage with the visualizations included in the atlases, but, similar to external experts, they may use a visualization atlas as a reference tool that can provide multiple resources on a topic to comprehensively understand its challenges.

\textbf{The General Public.}
5/7 atlases were designed with a ``general public'' audience in mind. Again, vaguely defined, this can include people without or with only little knowledge about the topic covered in the atlas, as well as people without expert knowledge in data analysis or visualization~\cite{burns2023we}. Some creators stated that they address people who have ``\textit{at least [seen] a chart before}''~[P2, P3; \textsc{SDG2023}] and that have an invested interest in the respective topic, e.g., because they are directly affected. For example, P7 framed the intended audience of \textsc{USClimate} as \textit{``people from the trenches on the ground level, [including] people engaged in local communities who care about climate issues deeply and inform funding proposals.''} \textsc{RareDisease} targets the parents and children who are directly affected by rare diseases.

\subsection{Design Practices \& Challenges}
\textbf{Design Inspirations.} Creators mentioned a range of inspirations that influenced their design, sometimes visual approaches, related projects, or broader practices from related communities. 
P8, for example, mentioned that they checked related atlas~\textsc{Opportunity}~\cite{OpportunityAtlas2023} or open data project like Trase~\cite{Trase2023}
\textit{``to see the new trends in the economic visualization world''}~[\textsc{EconComplexity}].
Other designers drew from science communication and data journalism in their design approach: \textit{``data journalism was the main inspiration, just because they [the visualizations] look really engaging. Whereas science communication is sometimes more dry looking.''}~[P2; \textsc{SDG2023}].

\textbf{Audience-Centred Design.} 
Most atlas projects target a range of diverse audiences (see above), and creators acknowledged this as a major design challenge. However, only 3/7 (\textsc{RareDisease}, \textsc{AIPolicy}, and \textsc{EconComplexity}) of the atlas projects in focus throughout our interviewed applied user-centered methods such as surveys, focus groups or interviews with target audience representatives as part of their design process. 
Some atlas creators acknowledged the lack of user-centered approaches as a potential problem (\textsc{SDG2023}), some highlighted intentions in studying the use and impact of their visualization atlases after it was finalized: ``\textit{over time, we can do user interviews [...], it is usually post projects}''~[P7, \textsc{USClimate}].

\textbf{Modes of Collaboration.}
Interviews show that designing visualization atlases is highly interdisciplinary; 6/7 atlases were created by interdisciplinary teams that included visualization designers, web developers, and domain experts. Only one atlas (\textsc{Budapest}) was created by a sole designer. 
\textit{Parallel Collaboration} was mentioned in the context of 3/7 atlases (\textsc{Calvino}, \textsc{EconComplexity}, \textsc{SDG2023}) with distinct roles assigned to domain experts and vis designers, in that domain experts would provide curated data alongside written stories, and the designers focused on the visual design and user experience.
\textit{Close Collaboration between Domain experts \& Vis Designers} was mentioned in the context of atlases(\textsc{AIPolicy}, \textsc{RareDisease}) where data was gradually sourced across the projects, and insight generation was still evolving during the design process.  Here, domain experts and designers worked closely together to decide (a)~what data and data stories should be included in the atlas and (b)~how to design visualizations and the atlas' general interface.
Our interviews indicate that this kind of collaboration allows for a gradual and experimental process of  atlas creation, where different versions are explored, before considering a final design that may then be expanded to a wider range of data or topics.

\subsection{Creators' Voices: What Makes a Visualization Atlas?}
The majority of projects covered in our interviews are named an \textit{``atlas''}~(\textsc{Calvino}, \textsc{SDG2023}, \textsc{Budapest}, \textsc{EconComplexity}), but other, related terms such as \textit{``Observatory''}~(\textsc{AIPolicy}, \textsc{RareDisease}), or \textit{``index''}~(\textsc{USClimate}) are also used. We therefore interviewed creators about the key aspects that, to them, make for a visualization atlas in comparison to other closely related approaches.

\textbf{Defined Topic $+$ Breadth of Perspectives \& Data Sources.}
Our interviews reveal that key to visualization atlases and comparable projects is the
focus on a \textit{\textbf{specific topic}}, presented in a data-driven way:
\textit{``I think you need a topic. [...] So the sustainable Development Goals served as our topic.''}~[P2; \textsc{SDG2023}]. \textit{``The main idea is to kind of collect all the data and show this overview for whoever is new [or] interested in the topic and wants to explore the details.''}~[P4; \textsc{AIPolicy}]. 
Creators also agreed that this topic should be approached from multiple perspectives and data sources:\textit{``You try to cover [the topic] throughout, not just like one angle, but different views of it [including coverage from research as well as the news].''}~[P4; \textsc{AIPolicy}]. 

\textbf{Topic \& Visual Curation.} 
Curation was highlighted as another key aspect of visualization atlases, concerning the topic, data and visuals: \textit{``You need a topic, you need a method of inclusion and exclusion, and you need a stylistic theme, that is coherent across the atlas.''}~[P2; \textsc{SDG2023}]. Topic and visual curation are linked: \textit{``every page helps you in understanding a topic or an aspect of a topic with the aid of visual materials.''}~[P1-\textsc{Calvino}].
5/8 creators acknowledged \textit{\textbf{data curation}}, i.e., deciding the in- or exclusion of data sources, as a key part of the atlas design process: \textit{``We had lots of discussions and trial-and-error getting the data, looking at it and [deciding] if we are missing something here, or if this is too much to present in a good way.''}~[P5; \textsc{RareDisease}]. The need to balance the granularity of data was also highlighted: \textit{``It's great to look at the broad trends, but can we get to more granular [perspectives]? So this broad view and then a specific area profile view marry each other very well.''}~[P7; \textsc{USClimate}].

Topical curatorial decisions can be political, reflecting on the authoring organization's identity and/or intentions. 
\textit{``We decided against [showing all of the targets in SDGs], because the World Bank is not the organization defining those targets. So they didn't want to overemphasize the whole structure [which] was more of a UN responsibility.''}~[P2; \textsc{SDG2023}]. At the same time, creators highlighted the need for transparency regarding curatorial decisions and data limitations because \textit{``for a traditional atlas, you give your trust to the cartographer or the statistical bureau that everything is 100\%. [...In] the data cut from international databases for Hungary, you can see some imperfections, inconsistencies in the data with our local knowledge''}~[P6; \textsc{Budapest}].

Our interviews further indicate that atlases demand \textbf{visual curation} to ensure visual consistency across the different sub-topics that may be presented within an atlas, which affects the design of corresponding visualizations: \textit{``you should design a sort of a graphical language that is reoccurring through the entire Atlas and with like color palettes and fonts and margins, layouts, grids and those kinds of things. Because you should be able to recognize that different visualizations are belonging to the same project.''}~[P1; \textsc{Calvino}]. 

\textbf{Debates on Comprehensiveness.}
An aspect that creators debated was the comprehensiveness of topic coverage. Creators stated that \textit{``you need to have a wide range of data, [...]~it has to cover the whole width of one specific aspect that is being worked on.''}~[P5; \textsc{RareDisease}]. However, some creators argued that atlas topics can be so big that comprehensiveness is not achievable. For example, P2 stated that the \textsc{SDG2023} is \textit{``less comprehensive in terms of how each goal is covered. I was surprised by that in the beginning, because I think my notion of an ``atlas'' would have meant that we are talking about every single target, [...] I understand why this is not feasible, because then you cannot go in depth enough at the same time. So they [the World Bank] wanted some depth in one aspect of each goal, rather than showing very surface-level insights.''}~[P2; \textsc{SDG2023}].

\textbf{Global \& Geospatial Perspectives.}
There was quite some debate about the necessity of a global, geospatial perspective as a defining element of visualization atlases:\textit{``I take the atlas in a very literal way, defining a geographic coverage that could be any kind of data encoding on top of it, like we map the economic data.''} [P8; \textsc{EconComplexity}]. However, other creators questioned the global nature of visualization atlases as a necessity: \textsc{USClimate} \textit{``is a national level data set. It doesn't matter which country that you have, but different levels of geographic granularity''}~[P7]. 
P6 questioned the necessity of geospatial aspects: \textit{``You want to see and understand visually what's happening, and it's not related only to maps or locations.''}~[\textsc{Budapest}].

\textbf{Up-to-Date Information.}
Creators further discussed the necessity for keeping visualization atlases up-to-date, which is linked to the requirement of working with dynamic data sources. For example, a new edition of the Atlas of Sustainable Goals is released every couple of years, often with a new set of targets in focus. Therefore, previous issues are not updated but show a snapshot in time. In contrast, presenting up-to-date information is a focus of \textsc{RareDisease} and \textsc{AIPolicy}. We found that naming conventions in the space of atlas-related projects matter. For example, the creators of \textsc{SDG2023} highlighted the term `atlas' \textit{``emphasizes the comprehensiveness [while] I think by calling it `observatory', they want to draw attention to the dynamic updates''}~[P2; \textsc{SDG2023}]. This was confirmed by the creators of \textsc{AIPolicy} and \textsc{RareDisease}: \textit{``I would like to think about an `observatory' as [something] having multiple streams of data, but also data gets updated. So it's about [...] observing through time.''}~[P4; \textsc{AIPolicy}].

\textbf{The Role of Interactivity.}
The importance of interactivity came across in 5/7 atlas projects mentioned in interviews. Although interactivity was not a key aspect in their design, the creators of the \textsc{SDG2023}, for example, mentioned that they included interactive elements in the visualization atlas to separate it from previous print-based issues. In other cases the support of interactive exploration is a key aspect: \textit{``Those visualizations really need to be interactive in order to make sense, [...], when you look at that feasibility chart, there are just so many nodes that it doesn't make sense without tooltip.''}~[P8; \textsc{EconComplexity}.
~Similarly, for the \textsc{USClimate} the \textit{``engagement factor is crucial. That ability to click around and drive it myself.''}~[P7].

\textbf{Comparison to Other Forms of Visualization.} The interviews also brought up debates about how visualization atlases relate to other types or forms of visualizations because of their difference in monitoring or storytelling purposes. The designer for \textsc{RareDisease} argued against infographics \textit{``because the idea is to to to have it like a exploratory tool''}, and she acknowledged an atlas can be made of \textit{``several dashboards put together with some storytelling''}. Yet P1 shared the opposite opinion that \textit{``the idea of narration is something that is very important to have in a project that is called Atlas. Otherwise, for me it's just a dashboard to compare, combine and visualize complex data. But it doesn't convey you the story''}. Similar ideas came across from the creators of \textsc{SDG2023} that \textit{``A dashboard is meant to monitor, usually updating data. So we don't have that [in the \textsc{SDG2023}]''}~[P2].

%% file: sections/genres.tex
\section{Visualization Atlases Genres}
\label{section:genres}

The design patterns in \autoref{section:survey} provide low-level design solutions and common practices while our interviews provide very detailed information about the rationales behind atlas design. In search of a practical middle ground, we now describe atlas \textit{genres}. Inspired by previous work on genres in narrative visualization~\cite{segel2010narrative} and dashboards~\cite{bach2022dashboard}, those genres can serve as templates for specific purposes and scenarios while informing the design decisions such as the choice of design patterns.

Informed by previous methodologies~\cite{segel2010narrative,bach2022dashboard,lan2023affective}, we searched for similarities between atlases based on their design dimensions. In this process and inspired by our interviews, we added one new dimension to the atlas coding---\textbf{data}---which was not part of our visual design pattern analysis. The data dimension captures if
\textit{(a)} data in an atlas are \textit{updated} or \textit{static}, 
\textit{(b)} if data come from \textit{internal} or \textit{external} sources, and 
\textit{(c)}~if an atlas is based on a \textit{single} or \textit{multiple datasets}. 

We used Bertifier~\cite{perin2014revisiting} and its table ordering techniques~\cite{behrisch2016matrix} to order rows (atlases) in \cref{fig:coding} by similarity of patterns across all dimensions, resulting in three visual clusters. Then we integrate interview findings in terms of motivations and topics to refine the clusters into 5 genres:
Storybanks, 
Observatories, 
Series,
Monographs, and Exploratoria.

\textbf{Genre: Storybank} aims to promote a defined, yet broad topic in a systematic, data-driven way. They are driven by multiple data sources, often external, that are updated regularly. Updates to a storybank may include the expansion of the topic and subtopics covered. Key atlas examples include \textsc{OurWorldInData} and \textsc{DataUSA}.

From a design perspective, storybanks are often defined by a large collection of \textit{articles} or \textit{reports} followed by an extendable structure of \textit{hierarchy}/\textit{multiple facets} plus \textit{parallel} pages. In \textsc{OurWorldInData}, the topic hierarchy creates placeholders for inserting new articles, insights, or charts. Multiple entry points are provided, including \textit{featured content} to enable visitors to pick-up recent highlights. However, storybanks have limited onboarding features, which, combined with their breadth and amount of content, can make the navigation more difficult.

\textbf{Genre: Observatory} shares similar design features as storybanks, but the topics are more narrowly defined and systematically tracked based on key performance indicators. The \textsc{AIPolicy} is a key example, where indicators are sourced externally,  processed, and presented together by-country and -topic to track the latest updates in AI policy. 

\textbf{Genre: Series} are published in more or less regular intervals. The visualization atlas series focuses on a high-level theme and each issue will not be updated systematically. Instead each issue features a selected (range of) topic(s) under that overarching theme, usually pertinent to its specific time periods. The topics are driven by internal data sources that frame in-depth stories, usually with a clear message. 
From a design perspective, atlas issues feature a simple, \textit{parallel} structure, consistent page style and different visualizations. However, the overall style, featured pages can be unique to each atlas issue and evolve over time. The \textsc{SDG2023} is the only example of a visualization atlas issue we discovered, belonging to \textsc{SDG series} issued by the World Bank. Over time, the series has featured more visualizations with more interactivity.

\textbf{Genre: Monograph} is a visualization atlas focusing on a highly specific topic with the purpose of promoting these for exploration and taking action. The corresponding data sources can be quite unique to the topic. Monographs are typically finished pieces without regular updates. As such, visualization atlas monographs are typically less expansive with their focus in depth than breath, featuring detailed stories and insights. From a design perspective, atlas monographs can follow tailored overall designs and highly customized visualizations. Key examples include the \textsc{Calvino} and the \textsc{Budapest} atlas.

\textbf{Genre: Exploratorium} promotes the active and open-ended exploration of particular topics by making extensive use of interactivity. Our exploratoria examples (such as the \textsc{EconComplexity}) are driven by internal datasets that are maintained and updated regularly by the author organization. Exploratoria can have quite complicated designs: they typically feature at least one \textit{exploratory vis} page as the core exploration tool, sometimes, but not always with supporting pages of either \textit{articles} or \textit{reports} to explain different perspectives on the topic in \textit{parallel}. There are also often multiple entry points and navigation strategies which are communicated through a range of onboarding strategies. 

%% file: sections/characteristics.tex
\section{Key Characteristics of Visualization Atlases}
\label{sec:characteristics}

In this section, we review our findings in detail and unpick 9 key characteristics grouped into three high-level aspects: \textsc{Topics}, \textsc{Curation}, and \textsc{Visualization} that best describe the nature of visualization atlases. 
The characteristics then inform a condensed and practical definition that a visualization atlas is \textit{a compendium of (web) pages aimed at explaining and supporting exploration of data about a dedicated topic through data, visualizations and narration}. To examine whether a project qualifies as a visualization atlas, the condensed definition captures the purpose and context of visualization atlases while the characteristics we describe in this section are specific criteria for inclusion and exclusion of what constitutes a visualization atlas.

\subsection*{\textsc{Topics}---Visualization atlases present complex topics, targeting a wide range of audiences. Topics are \textit{complex}, \textit{data-driven} and \textit{comprehensive}.}

\textbf{Complex}---Atlas topics are often of wide interest and target diverse audiences that include domain experts, either internal or external to the project, for research and further analysis, audiences in policy for general information and dissemination (rather than directly for decision making), as well as members of the general public including those in ``the trenches'': community workers and those directly concerned by the topic of the atlas. As such, the usage scenarios of atlases are very broad,
including promoting important global topics in a data-driven way,  promoting these data sets for exploration and action, and publishing insights and stories about both the data and topics. They aim to combine breadth and depth (horizontal and vertical exploration~\cite{dork2011information}).

\textbf{Data-driven}---Visualization atlases present topics through data. They
report and analyze typically multiple data sources around the topic. More specifically, they report on selected findings from these data, in the form of visualization-driven stories, reports or more exploratory interactive visualizations. By drawing from and carefully selecting data, visualization atlases promote the topic and enhance understanding of the available data---or the eventual absence of data. 

\textbf{Comprehensive}---Atlases aim to cover the respective topic in a meaningful and well-informed breath and depth. For example, \textsc{USClimate} covers the entire United States on the level of tracts, counties, states and EPA regions, reporting climate vulnerability from two groups of indicators: climate impact and community baseline. Likewise, \textsc{SDG2023} covers all 17 sustainable development goals.

\subsection*{\textsc{Curation}---Visualization atlases are highly curated in terms of content, data and presentation; they are \textit{scoped}, \textit{structured}, and \textit{visually curated}.}
The choices about topics and data require deliberate curation throughout the entire process of the atlas creation. Additional choices are required in combination with the targeted range of audiences and intended usage scenarios.
To this end, the author organizations---global NGOs, large or small research institutes, or design agencies---often bring together interdisciplinary teams with expertise in the topic, data, visualization and user interface design.

\textbf{Scoped}---Visualization atlases are highly scoped rather than being open repositories for arbitrary kinds of data.
Content curation concerns the framing of the topic, the choice of data, and the framing of stories. For example, in some atlases, certain data and related stories are deliberately excluded to maintain scope, to keep the content accessible to large audiences, but also to reflect on the nature, responsibilities and agendas of the organizations involved. The transparency and nature of these decisions can vary. In visualization atlas projects topic and data curation involve a handful of internal domain experts (top-down), whereas others involve the whole project team (including visualization designers) or even inform curatorial decisions through user-centred research (bottom-up).  

\textbf{Structured}---Atlases are structured collections of (web) pages with clear purposes and delineation for each page. Atlases usually provide entry pages, cross-links between pages, clear recurring (parallel) page structures, and aids for navigation and search. Structure aims to represent the topic (e.g., subtopics) and help understanding. Onboarding mechanisms help introduce those new to the topic and atlas.

\textbf{Visually curated}---Visual curation concerns how the topic, data and stories are visually (re-)presented and made explorable. This includes a sense of visual consistency across the atlas, while also recognizing the individuality of sub-topics and related visualizations and pages. 
It also includes establishing a coherent visual identity across pages, clearly communicating navigation, and facilitating user onboarding. Visual identity can be created through using consistent and unique (combinations of) colors, font types, page layout, visualization style, etc. A visual identity is important to make the atlas a coherent piece and to support its authority. This also applies to general accessibility to meet the needs of readers with bodily impairments, responsiveness to different device contexts, and audiences with low levels of general literacy, data literacy, numeracy, or visualization literacy. 

This high degree of curation ensures a focused engagement with the specific topic of the visualization atlas. Without the criteria of \textsc{curation} and \textsc{topic}, atlases would be more like other repositories, or collections of data and visualizations. Naturally, curation processes are heavily influenced by the creator organizations, their profiles, and intentions. We will discuss this further in \cref{sec:discussion}.

\subsection*{\textsc{Visualization}---Visualization atlases heavily rely on data visualization; they are \textit{visualization-driven}, \textit{explanatory} and \textit{exploratory}.}

\textbf{Visualization-driven}---Data visualizations serve as the main medium to engage with the topic and data presented in a visualization atlas. 
This characteristic aligns well with open data initiatives that improve public engagement through open data and data visualizations such as discussion platforms~\cite{viegas2007manyeyes} and public discourse~\cite{knudsen2018democratizing}.
Visualization atlases are not bound to geospatial visualization, but can incorporate a variety of charts and related visualization techniques, sometimes even quite non-standard (\textsc{Calvino}). The use of consistent visualization templates can emphasize the holistic nature of the visualization atlas, while individualized visualization designs can emphasize the nature of specific sub-topics. Importantly, visualizations in atlases can follow both reader-driven and author-driven approaches~\cite{segel2010narrative}; they  are both exploratory and explanatory:

\textbf{Explanatory}---Visualizations in atlases serve explanation through e.g., visual curation, annotations, textual explanations. Visualizations for storytelling are integrated into articles and reports to communicate and emphasize stories and insights (author-driven). They complement text and annotations through visual access to the data; they can re-iterate the information in the text, or serve as an additional information source.

\textbf{Exploratory}---Visualizations also serve exploration through being interactive, providing detailed and complementary views of the data. They may attract curiosity and function as illustrations on entry pages to communicate the overall mission and topic of an atlas. Those pages that allow for more flexible, free-form exploration may take the style of exploratory visualization, dashboards, and chart collections that are rich in interaction (reader-driven).

%% file: sections/discussion.tex
\section{Discussion}
\label{sec:discussion}

\textbf{Uniqueness of visualization atlases.} Our definition and its key characteristics helps us discuss visualization atlases' uniqueness with respect to other forms of visualization and collections.  
Visualization atlases are higher-level structures that incorporate other genres and forms of visualizations in their content pages: analytic interfaces (interactive visualizations), dashboards~\cite{bach2022dashboard} or narrative visualizations~\cite{segel2010narrative} such as articles~\cite{hao2024design}, comics~\cite{bach2018design}, videos~\cite{amini2015understanding}, etc. 
Some atlases use a limited number of content page design patterns, while others include a wider range and mix components from news articles, dashboards, and slide-shows. 
Atlases are also different from mere collections of articles (e.g., blogs, news websites) as these collections usually lack framed \textit{structure} of those pages, making each page a rather independent contribution without \textit{scoped} content curation.
Likewise, collections of visualizations like VISAP or survey repositories are typically not \textit{data-driven} but rather compilations of visualizations that are neither \textit{explanatory} nor \textit{exploratory} in the context of the survey.
Dashboards are an interesting case since they share many of an atlas' characteristics; \textit{data-driven, scoped, structured, visually-curated}, and \textit{visualization-driven}. 
However, dashboards focus on monitoring and at-a-glance-overviews rather than supporting explanation or being comprehensive, especially since they usually lack mechanisms for onboarding and narration. Instead, atlases can be made of pages of dashboards, one of the content page types.
Finally, atlases are different from encyclopedias (e.g., Wikipedia) or online galleries (Google Arts \& Culture, Pinterest), as these are not \textit{data-driven} nor \textit{visualization-driven} within a topic.

\textbf{Limitations of methods and definitions.} 
Visualization atlases are of considerable complexity and richness. 
Our definition for visualization atlases is derived from shared qualities in the survey collection. This paper aims to describe these qualities and understand visualization atlases in the context of emphasizing notions of complex topics, accessibility and structure. 
The boundaries between the qualities and notions can be fuzzy as each atlas case encapsulates human ideas and inventions. 
Our identified genres (\cref{section:genres}) capture groups of common qualities and usage contexts.
We took an explorable approach in collecting atlas cases, meaning that because of the lack of a common terminology or concept of visualization atlases, we might have missed similar projects in nature which use different names. However, we believe our definitions and characteristics describe this emerging class of projects sufficiently well to justify the existence of this new format.

\textbf{Extending atlases.}
We did not encounter visualization atlases incorporating visualizations such as data comics~\cite{bach2018design, wang2021interactive}, data-gifs~\cite{shu2020makes}, or data videos~\cite{amini2015understanding, shen2023data}. These and other forms/genres of data representation (e.g., sonification or data physicalization~\cite{Jansen_2015}), as well as related approaches (e.g., interactive notebooks, explainable explorables~\cite{explainableexplorables}, immersive technology~\cite{isenberg2018immersive}, or streaming~\cite{zhao2022datatv}) offer a vast range of design opportunities for visualization atlases. All these building blocks alongside our 33~examples can inform the design of future atlas projects, especially those involving teams. They can inform design card decks~\cite{roy2019card,bach2018design}, design activities~\cite{huron2020ieee}. Future research could look into specific toolkits and web-design platforms for making atlas design, development, and maintenance easier. We hope this paper inspires future research to imagine new genres of visualization atlases that may also expand the design dimensions and patterns presented here.

\textbf{Designing visualization atlases in a user-centred way.}
Our interviews have highlighted that visualization atlases potentially speak to a wide range of audiences, including researchers, domain experts, educators, policy makers, and the `general public'. This imposes several design challenges. Some of these audience groups are notoriously difficult to define and characterize, which, in turn, makes it difficult to define usage scenarios and do justice to the diverse goals and needs of different audiences. This echoes recent research~\cite{he2023enthusiastic} that highlights the existence of varying levels of audiences' openness to public data and visualizations.
Given these challenges and the significant real-world impact of visualization atlases as tools of authority, the integration of user-centered methods in the design process can more effectively address specific usage scenarios and audience perceptions. While our interviews indicate that creators are aware of this issue, only 3/7 of the atlas projects in our interview actually applied user research due to limited resources, and more bottom-up approaches such as co- or participatory design were not mentioned at all. 

We therefore argue for future design practices in the context of visualization atlases to put a stronger emphasis on allowing 
\textit{a)} target audiences and 
\textit{b)} people who are at the center of or potentially affected by the presented topics and data to actively participate in atlas design processes (as early as the stage of data creation/selection and curation). How this can be achieved in practice, given the complexity of visualization atlases that often already involve large interdisciplinary teams, and how to best document and share best practices is an interesting challenge but also an opportunity for future research.

\textbf{Real-world impact of visualization atlases.}
Visualization atlases potentially have a huge real-world impact, due to the fact that they often focus on topics of global relevance and because atlases are deliberately designed to advance access to these topics. At the same time, a lot of visualization atlas projects are initiated by impactful and large organizations which give these atlases reach and credibility.
Besides educating us about certain topics, atlases may promote a culture of data-driven evidence and decision-making and increase data and visualization literacy. However, the use and impact of visualization atlases in different contexts and on different audiences is largely unexplored. The creators we interviewed acknowledged that this is usually not explored as part of the atlas project. Future research in this area is crucial to 
\textit{a)} inform visualization atlas design decisions, and 
\textit{b)} explore their potential impact on research, education, policymaking, etc.
Respective tools to investigate the real-world impact of visualization atlases could include a range of qualitative research methods such as interviews and observations. However, we may also draw from theory such as feminism (e.g.,~\cite{risam202335,Ignazio_2016}), or critical perspectives on visualization~\cite{Doerk_2013, Correll_2019} to interrogate the context and design of existing visualization atlases and the power structures into which they potentially feed.

\textbf{Transparency of curation processes.}
As we have highlighted in this paper, curation processes are a key characteristic of visualization atlases. Content and visual curation ensures consistency and coherence in ways a visualization atlas presents and discusses the topic in focus. However, given the fact that visualization atlases potentially have a big real-world impact, making these curation processes transparent is of utmost importance and a huge challenge at the same time. Transparency not only concerns data provenance~\cite{burns2022invisible}, i.e., the origin of data sources and the data transformation processes that preceded the visualization and narratives included in the atlas. It also concerns transparency about the goals and intentions behind the atlas, including the organizations involved and their (political) profile, the atlas team, also acknowledging the labor of external people, and the decision-making processes that influenced the selection of topic(s), inclusion/exclusion of data (sources), design processes, etc. Interviews with creators show an awareness of the importance of transparency, but also highlight that it remains a challenge (\textsc{SDG2023}). 

Highlighting the limitations of an atlas can be quite important in this regard. Recent work on provenance in visualization~\cite{burns2022invisible,vancisin2023provenance}, uncertainty visualization~\cite{padilla2020uncertainty,Panagiotidou_2023}, and visualizing missingness in data~\cite{Fernstad_2019,Song_2019} may be a starting point here. Correll~\cite{Correll_2019} has reminded us of the ethical dimensions of data visualization and of our responsibilities as visualization researchers and designers. In this context of visualization atlases which we can consider tools of power, just like their traditional eponyms, establishing rigorous, reliable and ethical methods is an important and interesting challenge.

\section{Conclusion}

This is the first paper to describe visualization atlases and to propose a definition for this emerging genre. We investigated 33 atlases from different angles, yielding design patterns, genres, and first-hand reflection from atlas designers. We believe visualization atlases are ``here to stay'' as they present solutions for how we can deal with large, complex, and urgent global topics understood through data---providing access to data, promoting knowledge, educating people, supporting further analyses, make data open. We believe visualization atlases are crucial for a data-literate society and a culture of data-driven (inclusive) decision making, balancing access to data and critical engagement.